\documentclass[9pt,twocolumn,twoside]{osajnl}

\journal{ao} 

\setboolean{shortarticle}{false} 
\usepackage{graphicx}

\ifthenelse{\boolean{shortarticle}}{\colorlet{color2}{color2b}}{\colorlet{color2}{color2}} 

\title{Closed-form Harmonic Contrast Control with Surface Impedance Coatings for Conductive Objects}

\author[1,*]{Giuseppe Labate}
\author[2]{Symon Podilchak}
\author[3]{Ladislau Matekovits}

\affil[1]{Department of Electronics and Telecomunications, Politecnico di Torino, Corso Duca degli Abruzzi 24, I-10129 Torino, Italy.}
\affil[2]{Institute of Sensors, Signals and Systems, Heriot-Watt University, Edinburgh Campus, EH14 4AS, United Kingdom.}
\affil[3]{Department of Electronics and Telecomunications,
Politecnico di Torino, Corso Duca degli Abruzzi 24, I-10129
Torino, Italy, and with the Macquarie University, Sydney, NSW Australia.}

\affil[*]{Corresponding author: giuseppe.labate@polito.it}

\dates{Compiled \today}

\ociscodes{(230.3205) Invisibility cloaks; 
(290.5839) Scattering, invisibility;
(290.0290) Scattering;
(160.3918) Metamaterials;}

\doi{\url{http://dx.doi.org/10.1364/ao.XX.XXXXXX}}

\begin{abstract}
The problem of suppressing the scattering from conductive objects is addressed in terms of harmonic contrast reduction. A unique compact closed-form solution for a surface impedance $Z_s(m,kr)$ is found in a straightforward manner  and without any approximation as a function of the harmonic  index $m$ (scattering mode to suppress) and of the frequency regime $kr$ (product of wavenumber $k$ and radius $r$ of the cloaked system) at any frequency regime.
In the quasi-static limit, mantle cloaking is obtained as a particular case for $kr \ll 1$ and $m=0$. In addition, beyond quasi-static regime, impedance coatings for a selected dominant harmonic wave can be designed with proper dispersive behaviour, resulting in improved reduction levels and harmonic filtering capability.
\end{abstract}

\setboolean{displaycopyright}{true}

\begin{document}

\maketitle
\thispagestyle{fancy}

\ifthenelse{\boolean{shortarticle}}{\ifthenelse{\boolean{singlecolumn}}{\abscontentformatted}{\abscontent}}{}


The challenge of creating devices for electromagnetic invisibility and cloaking has fascinated the scientific community with increasing research efforts: from finding plasmonic or epsilon-near-zero covers \cite{PC, PC2} and metamaterial shells for controlling light with geometry \cite{TO} to the use of scalar and high order impedance models for patterned metasurfaces \cite{MC,PattMeta,Jiang_Werner}. 
The robust mathematical theory based on Transformation Optics \cite{TO} has been derived considering the coordinate transformation of a background scenario (by construction, not perturbing fields) into another curvilinear  coordinate system supporting the same non-scattering phenomenon, thus finding unexpected reflectionless systems. For the harmonic suppression of a general scattering event, Scattering Cancellation has been developed considering plasmonic cloaking with bulk metamaterials \cite{PC, PC2} and mantle cloaking with thin metasurfaces \cite{MC, PattMeta}.
\begin{figure}[t]
\centering
\includegraphics[width=0.48\textwidth]{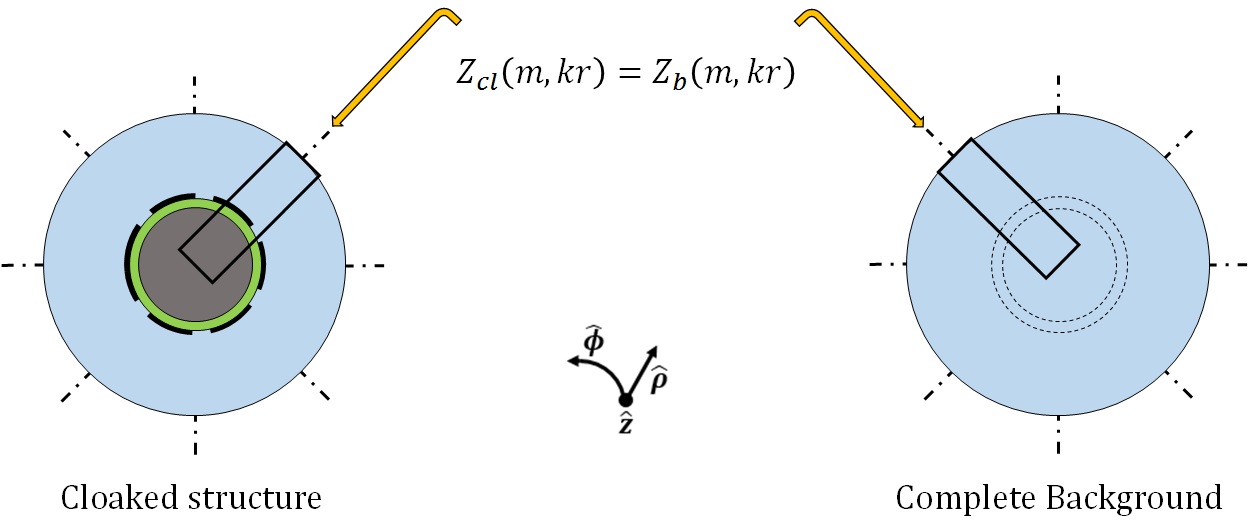}
\caption{\small Cylindrical geometry (top view) with attached coordinate system (center): metasurface wrapped around a conductive object (left) and complete background system (right). }
\label{fig:setup_prob}
\end{figure}
To this purpose, Mie Theory \cite{Born_Wolf} has been used for setting the problem of finding the zeros of the external scattering coefficients for  dielectrics and conductive objects at optical frequencies \cite{PC2,Cui}.
In this respect, closed-form equations have been established for dielectric cloaking in the quasi-static regime \cite{MC,PattMeta}, due to the fact that it is possible to cloak a dielectric object of radius $a$ with an ultrathin impedance metasurface at $r=b$ directly attached to the object itself, thus $b=a$. In this simplified scenario, a compact closed-form solution has been found and generalized  at any frequency regime for dielectrics quite recently \cite{Lab_alu_Mat}. For conductive objects, the zeros of the external scattering can only be controlled at a distance $r=b>a$, detached from the object itself, and it remains challenging and cumbersome, as stated in \cite{Monti} for example, even to find approximate formulas in quasi-static regime for opaque/impenetrable objects. This appears to be more and more complicated when the device is beyond the quasi-static regime. 

In this work, we apply a general methodology for reducing the scattering, as first introduced in \cite{LabyOPEX} and inspired by the \textit{contrast} concept as applied here to conductive objects: since in optical vision applications, the final goal for visibility is the contrast enhancement of harmonic scattered waves \cite{contrast}, a proper reduction of the contrast leads to zero and low harmonic scattered waves.

In the following treatment, we will consider 2D cylindrical geometries, but the method can be in a straightforward way adapted to 3D structures when spherical waves  are considered \cite{Dana_apop}. A closed-form solution for the dominant harmonic cancellation has been obtained when an impedance coating is directly attached to the surface of a dielectric cylinder \cite{Lab_alu_Mat}. A more general design equation is obtained in this work, considering that for conductive objects the impedance cloak can not be directly attached to the surface itself, but it has to be at a certain distance, even if the spacer is thin \cite{Sounas}.
As illustrated in detail in Fig. \ref{fig:setup_prob}, a conformal metasurface, made up of a ground plane, dielectric spacer and impedance loading (left) can behave for external observers as the same geometry but completely filled by background  material (right). If the contrast between these two structures is set to zero for a certain parameter (here, the input impedance), the cloaked structure will behave similarly to a complete background scenario. 

We compute the impedance to a cylindrical wave incident on a complete background scenario (ideal solution of any cloaking problem) under TM$_z$ polarization (electric field parallel to the cylinder's axis), by considering plane waves as expandend in $m$-harmonic cylindrical waves 
\begin{align}
& E^z_i(k_b r,\phi)= \sum_{m=-\infty}^{m=+\infty} i^m\ \mbox{e}^z_i(m, k_b r)\ e^{-i m\phi}  \\
& H^\phi_{i}(k_b r,\phi)= \sum_{m=-\infty}^{m=+\infty}i^m\ \mbox{h}^\phi_{i}(m,k_b r) \ e^{-i m\phi}
\end{align}
where $k_b$ is the background wavenumber and the radial basis functions are
\begin{align}
\label{eq:ei_field}
& e^z_i(m,kr)=J_m(k_b r) \\
\label{eq:hi_field}
& h^{\phi}_i(m,kr)= -i Y J'_m(k_b r)
\end{align}
where $Y$ is the admittance as computed in the $\phi z$-plane at the same radial distance $r$, whereas $J_m(\cdot)$ and  $J'_m(\cdot)$ are the Bessel function of order $m$ and its derivative, respectively. According to \eqref{eq:hi_field}, a time dependence $e^{i\omega t}$ is tacitly assumed throughout this work.

Even if the background is free of source and objects, the  value $\widetilde{Y}_{b}$, normalized with respect to the background intrinsic admittance $Y_B=\sqrt{\varepsilon_b/\mu_b}$, is obtained as %
\begin{align}
\widetilde{Y}_b\equiv \dfrac{1}{Y_B} \dfrac{h^{\phi}_i(m,kr)}{e^{z}_i(m,kr)} =  -i\dfrac{\mathrm{J}'_{m}(k_b r)}{\mathrm{J}_{m}(k_b r)}
\label{eq:y_b}
\end{align}
and it represents the input admittance of free-space for each single cylindrical harmonic wave as in \cite{Lab_alu_Mat}.
 
For the cloaked system, the case study will be reducing the scattering from a conductive cylinder by obtaining a proper surface impedance function $Z_s(m,kr)$, where the dependence on $kr$ indicates the dispersive behaviour and the harmonic index $m$ the mode to be suppressed.

For a cylindrical conductive object with a circular cross section, using the cylindrical coordinate system $(r,\phi,z)$,  the propagation of the (incident and scattered) energy occurs through a radial direction (the wavenumber is $\vec{k}\equiv|k|\hat{r}$) and, as a consequence, through cross sections $A$ that are similar, but non-uniform, because of the radial variance along the structure while traveling, forming $\mbox{d}A(r)=r \mbox{d}r \mbox{d}\phi$. 


Such complication is considered in Mie Theory by choosing  cylindrical waves modeled as Hankel functions of the first (inward) and second kind (outward) of order $m$. In the annular spacer $a\leq r \leq b$, the electric and magnetic fields are expanded as
\begin{align}
\label{eq:e_field}
& e^{z}_t(m,kr)=E^- H_m^{(1)}(kr)+E^+ H_m^{(2)}(kr) \\
\label{eq:h_field}
& h^{\phi}_{t}(m,kr)= -i Y \begin{bmatrix}E^- H_m^{(1)'}(kr)+E^+ H_m^{(2)'}(kr)\end{bmatrix}
\end{align}
where $k$ is the wavenumber in the ring filled with dielectric material, whereas $E^-$ and $E^+$ are the reflected and transmitted waves, respectively (magnetic fields are instead proportional to Hankel derivatives). In the external region $r \geq b$, the scattered electric and magnetic fields are expanded with radial functions as
\begin{align}
\label{eq:es_field}
& e^z_s(m,kr)=c_m H_m^{(2)}(kr) \\
\label{eq:hs_field}
& h^{\phi}_s(m,kr)= -i Y c_m H_m^{(2)'}(kr)
\end{align}
where $c_m$ is the scattering coefficient in the background region, weighting outward waves according to classical Mie Theory for non-scattering structures \cite{PC, MC, Cui}.

For cloaking conductive cylinders, two boundary conditions are here imposed considering total field functions: one at $r=a$, enforcing a zero impedance condition, and a second boundary condition that has to ensure a zero contrast admittance condition (thus, a cancellation effect, see Fig. \ref{fig:setup_prob}) at $r=b$, the external radius of the cloaked-background interface, as
\begin{align}
\label{eq:bc_1}
& Z(r=a)\equiv\dfrac{e^{z}_t(m,ka)}{h^\phi_t(m,ka)} =0 \\
\label{eq:bc_2}
& \begin{bmatrix}\dfrac{h^\phi_t(m,kb)}{e^z_t(m,kb)}  + Y_s\end{bmatrix}-\dfrac{h^\phi_i(m,k_b b)}{e^z_i(m,k_b b)}  =0 
\end{align}
for each specific harmonic mode $m$. 

Solving for the admittances,  the second boundary condition gives rise to an equation for the contrast between the input admittances $\widetilde{Y}_{cl}$ (cloaked cylinder) and $\widetilde{Y}_b$ (background cylinder).


%
%
%

The input admittance $\widetilde{Y}_{cl}$ (cloaked device) contains two contributions: a term $\widetilde{Y}_{d}$, obtained by looking at the conductive object through the dielectric ring and a second term $\widetilde{Y}_s$, the normalized surface admittance for loading the impedance metasurface at $r=b$ in order to tune the contrast to zero. \eqref{eq:bc_2} is equivalent to compute the zeros for the external scattering coefficients $c_m$ as in classical Mie Theory for cloaking \cite{PC,MC}, but in a straightforward manner and without any approximation as follows.

The reflection occurring at  $r=a$ is described by the ratio $\gamma(ka)$ between the reflected ($E^-$) and the transmitted ($E^+$) waves at $r=a$,  obtained from \eqref{eq:bc_1} as
\begin{align}
\gamma(ka)\equiv \dfrac{E^{-}}{E^{+}}=-\dfrac{\mathrm{H}_{m}^{(2)}(ka)}{\mathrm{H}_{m}^{(1)}(ka)}
\end{align}
It is worthwhile mentioning that now the wavevector,  imposed at the inner boundary by the presence of the dielectric spacer, is $k=\sqrt{\varepsilon_r}k_b$. As regards the boundary at $r=b$, dividing \eqref{eq:h_field} by \eqref{eq:e_field}, it is possible to obtain the admittance function through the radial direction as a function of the ratio $\gamma(ka)$ as
\begin{align}
& \widetilde{Y}_d(r=b)\equiv \dfrac{Y_d}{Y_B}=  - i \sqrt{\varepsilon_r}\begin{bmatrix}\dfrac{\mathrm{H}_{m}^{(2)'}(kb)+\gamma(ka)\mathrm{H}_{m}^{(1)'}(kb)}{\mathrm{H}_{m}^{(2)}(kb)+\gamma(ka)\mathrm{H}_{m}^{(1)}(kb)}\end{bmatrix}
\label{eq:y_d}
\end{align}
Since the input admittance $Y$ is seen from the free-space region, there is an amplified factor $\sqrt{\varepsilon_r}$ in front of \eqref{eq:y_d}, in order to pass from $r=b^-$ at $r=b^+$ \cite{Marcuvitz}, where also the surface admittance $Y_s$ is located in a parallel fashion (considered to lie above the dielectric ring, thus in free-space). In a straightforward manner, the unknown admittance $Y_s$  can be written in a compact form, substituting \eqref{eq:y_b} and \eqref{eq:y_d} in \eqref{eq:bc_2}, thus solving for $\widetilde{Y}_s\equiv Y_s/Y_B$  obtaining 
\begin{align}
\label{SACE}
& \widetilde{Y}_s=  
-i\dfrac{\mathrm{J}^{'}_{m}(k_b b)}{\mathrm{J}_{m}(k_b b)} +  i\sqrt{\varepsilon_r} \begin{bmatrix}\dfrac{\mathrm{H}_{m}^{(2)'}(kb)+\gamma(ka)\mathrm{H}_{m}^{(1)'}(kb)}{\mathrm{H}_{m}^{(2)}(kb)+\gamma(ka)\mathrm{H}_{m}^{(1)}(kb)}\end{bmatrix} 
\end{align}
where the denormalized $Y_s$ is the required surface admittance for the cancellation of the $m^{\mbox{th}}$-harmonic as a function of the three main parameters of the geometry under consideration: the radius of the conductor $a$, the relative dielectric permittivity of the spacer  $\varepsilon_r$ and the external radius $b$, where the admittance/impedance load is attached.
The obtained $Y_s$ formula for conductive objects in \eqref{SACE} shows two main advantages as the one for dielectrics \cite{Lab_alu_Mat}: it is directly given in closed-form with its dispersive behaviour (not constant across the frequency band) and it is selectively related to the $m$-harmonic index for specific scattering cancellation.
The surface admittance/impedance function depends not only on geometrical and physical parameters ($a,b,\varepsilon_r$), related to the specific cancellation of the harmonic ($m$), but also exhibits $k$-dispersion (or $\omega$-dispersion, where the angular frequency is $\omega=2\pi f$).

\begin{figure}[t!]
\centering
\includegraphics[width=0.4\textwidth]{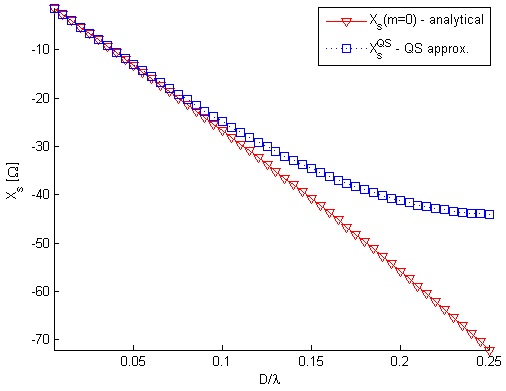}
\caption{\small Dispersion of $X_{s}$ as a function of $D/\lambda$ (linear scale): reactance for mantle cloaking \cite{Monti} $X_s^{QS})$ (blue square dots) and for exact analtical solution $X_s(m=0)$ (red triangle dots). }
\label{fig:Zs_0}
\end{figure}

The challenge of analytical formulas for $Z_s$ has been faced by mantle cloaking, but only  the quasi-static function $X_{s}^{QS}(\omega)$ is avalaible in the literature for a conductive cylinder  \cite{Monti},  here  reported for sake of comparison as
\begin{align}
\label{eq:Monti_cl}
X_{s}^{QS}(\omega)=- 16\mu_0\omega \dfrac{b(b^2k_0^2-4)\log(b/a)}{(a^2k_0^2-4)(256-4b^2-k_0^2+b^6k_0^6)}\\ \times [4(a^2-b^2)k_0^2+(a^2k_0^2-4)(b^2k_0^2-4)]
\nonumber
\end{align}
and plotted in Fig. \ref{fig:Zs_0}.
Authors in \cite{Monti} have found \eqref{eq:Monti_cl} for TM$_z$ case by expanding in Taylor's series Bessel and Hankel functions involved in the overall scattering process as in Mie Theory analysis \cite{Monti}. 

As reported for example in Fig. \ref{fig:Zs_0}, the surface impedance, needed for cancellation effects, appears to be purely reactive if $k_b$ and $k$ are wavenumbers propagating in media without losses. However, it was demonstrated by Foster \cite{Foster} that, no matter how complex a reactance network is, the dispersion behaviour of lossless impedance systems is always obeying to the law 
\begin{align}
\dfrac{\partial X_s(\omega)}{\partial \omega}>0
\end{align}
which states that pure imaginary $Z_s(\omega)=\jmath X_s(\omega)$ must monotonically increase as a function of the frequency. As reported in Fig. \ref{fig:Zs_0}, unfortunately this is not the case for $X_{s}(\omega,m=0)$, here intended as the surface impedance needed for the suppression of the zero order harmonic in TM$_z$ polarization: as a consequence, narrowband operation is expected for a single homogeneous surface impedance when passive coatings are used. However, this closed-form solution can be appealing for active metasurface cloaks \cite{Non_Fost} for any arbitrary dominant $m$ index to suppress, in order to go beyond the limit as imposed by all passive coatings \cite{Mont_alu}.

\begin{figure}[ht!]
\centering
\includegraphics[width=0.49\textwidth]{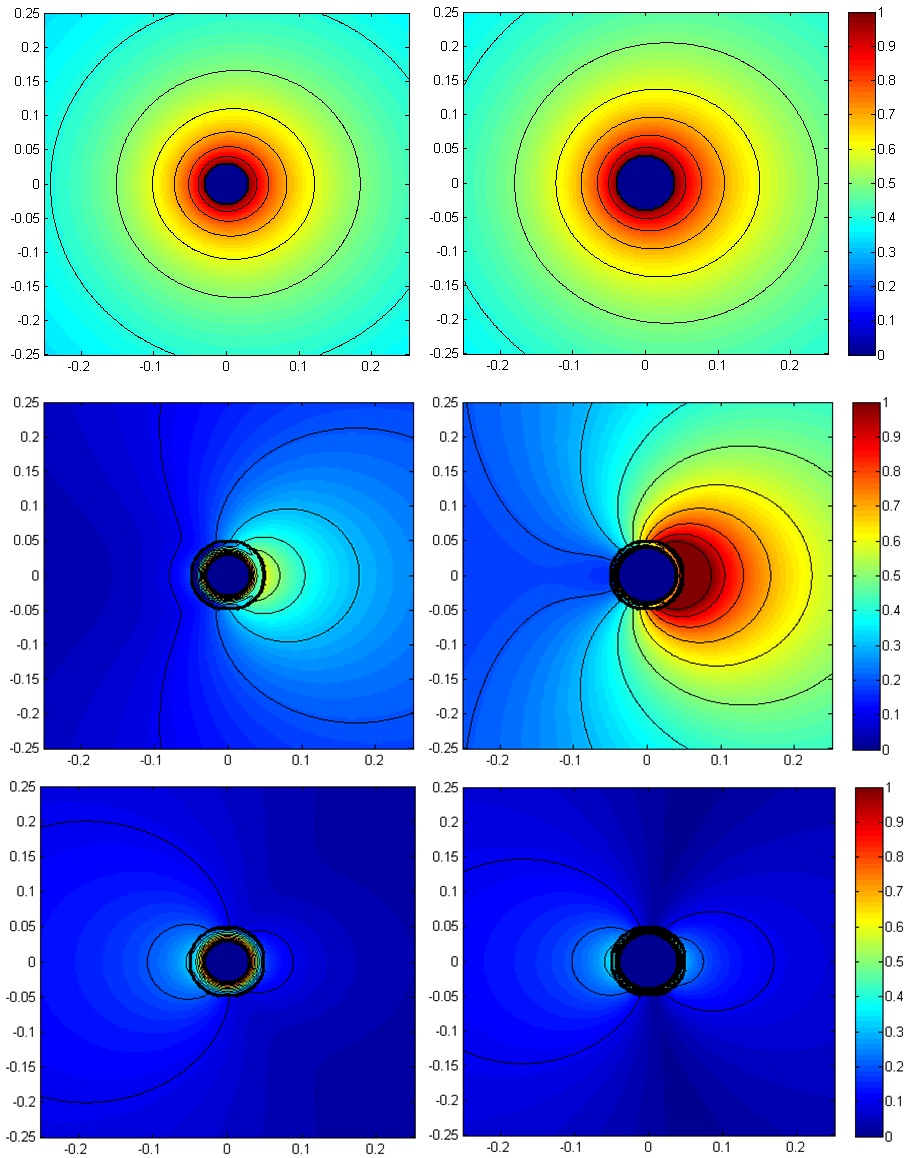}
\caption{\small Numerical results with Mie Theory \cite{Cui} for the absolute value (linear scale) of scattered field $E_s$ ($D\equiv 2b=0.1\lambda$) by changing  $a=0.03\lambda$ (left column) and $a=0.04\lambda$ (right column): uncloaked case (top), cloaked case with $X_s^{QS}$ (middle) and cloaked case with exact $X_s(m=0)$ (bottom). }
\label{fig:qs_exact_cloak}
\end{figure}

For practical implementation as addressed in this work, this closed-form design gives an alternative procedure, at any frequency regime, if a dominant mode cancellation is sufficient for global scattering reduction without complicating the metasurface cloak with surface resistance elements for balanced loss and gain \cite{Sounas}.

%
Even if in Fig. \ref{fig:Zs_0} both curves are quite superimposed in quasi-static regime up to $D\equiv 2b<0.1\lambda$, as the device starts increasing in terms of wavelength, the gap between $X_s^{QS}(k_b r \ll 1)$ and  $X_s(k_b r,m=0)$ starts increasing for the same physical and geometrical structure. In order to improve the cloaking performace with exact surface impedance solutions, the graph in Fig. \ref{fig:Zs_0} has been exploit as a design tool: the impedance coating has been computed for a conductive cylinder of two different radii  ($a=0.03\lambda$ and $a=0.04\lambda$), covered by an air spacer at a fixed radial distance $b=0.05\lambda$. For the bare conductive cylinder with $a=0.03\lambda$, the quasi-static formula gives $X_s^{QS}=-56.62\ \Omega$ with respect to the exact solution for the mode $m=0$ that results to be $X_{s}(m=0)=-62.52\ \Omega$. 
For the uncloaked case with $a=0.04\lambda$, the quasi-static formula gives $X_s^{QS}=-24.91\ \Omega$ whereas the exact solution for the mode $m=0$ gives $X_{s}(m=0)=-26.77\ \Omega$. 
The surface impedance values, obtained inverting \eqref{SACE} and from mantle cloaking via  \eqref{eq:Monti_cl}, as reported in Fig. \ref{fig:Zs_0}, are plotted as a function of the maximum dimension of the cloaking system in terms of wavelength (i.e., $D/\lambda=2b/\lambda$), where the cloak is placed at $b=1.25a$ with a dielectric layer made of air (i.e., $k=k_0$).
\begin{figure}[t!]
\centering
\includegraphics[width=0.45\textwidth]{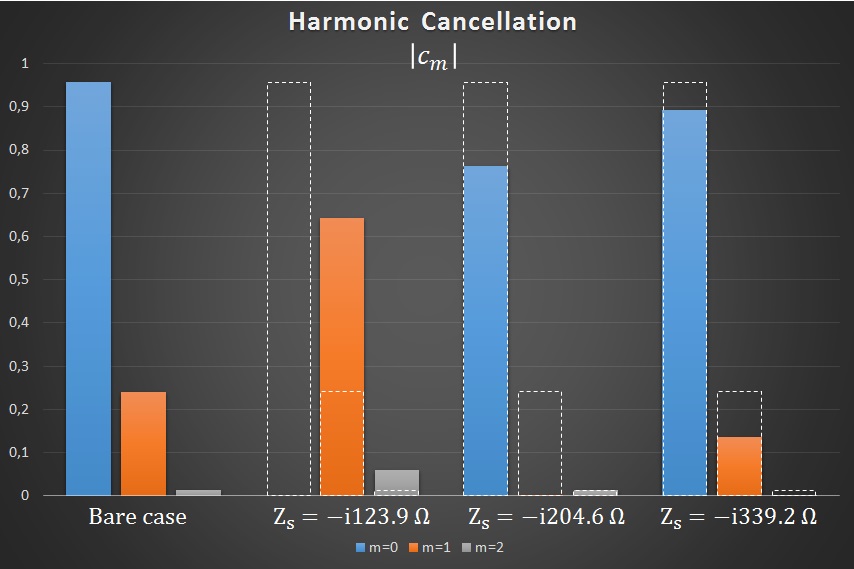}
\caption{\small Harmonic cancellation effect considering $|c_m|$, for $m=0,1,2$ with the computed impedance values: bare and cloaked case for $m=0$ ($Z_s=-i123.9\ \Omega$), for $m=1$ ($Z_s=-i204.6\ \Omega$) and $m=2$ ($Z_s=-i339.2\ \Omega$) according to the synthesis via \eqref{SACE}. Bare case is also reported in dashed-line for comparison.}
\label{fig:h_canc}
\end{figure}

In order to check the validity of the synthesis process in comparison with the quasi-static formula, a Mie Theory analysis is performed for cloaking conductive scatterers as detailed in \cite{Cui}. 
As reported in Fig. \ref{fig:qs_exact_cloak}, even if such case is at the  border of the quasi-static condition (external diameter fixed at $D= 0.1\lambda$) and thus the reactance value differs only of few $\Omega$ in this frequency regime, the absolute value of the scattered field is different for the two cloaked cases (quasi-static and exact formula) and also when the radius $a$ changes. With respect to the bare conductive cylinder (top), a residual scattering in the forward direction is always observed in the cloaked case according to the quasi-static formula (and even deteriorates when $a=0.04 \lambda$). However, when the exact solution is employed, a deep reduction in the outside region is always observed: as a comparison, contour field level are inserted and set for all three cases to be equal with the same step $\Delta|E_s|=0.1 $ V/m. 

In addition, beyond quasi-static regime, as reported in Fig. \ref{fig:h_canc}, the scattering from uncloaked and different cloaked structures is tested, in order to validate if \eqref{SACE} is able to give the exact value of the $Z_s(m)$ needed to cancel the selected harmonic wave. The reference uncloaked scenario is a bare conductive cylinder of radius $a=0.1\lambda$ and all the cloaked devices are considered with a low-contrast dielectric $\varepsilon_r\approx 1$ but loaded with different impedance values at $b=0.15\lambda$ (thus, looking at Fig. \ref{fig:Zs_0}, beyond the quasi-static regime with $D/\lambda = 0.3$). 

The first three cylindrical harmonics with $m=0,1,2$ are given as a function of the absolute value of $c_m$, detected by external observers \cite{PC,MC,Cui}. For this specific uncloaked case, the dominant term is still $m=0$ and this harmonic is set to zero by the impedance value $Z_s=-i 123.9\ \Omega$. However, there is an increase in the field level for the two successive harmonics $m=1$ and $m=2$ with respect to the bare case. The selective filtering of the impedance metasurface is evident when the other cylindrical hamonic waves are chosen to be suppressed via  \eqref{SACE}. For the specific value $Z_s=-i204.6\  \Omega$, the harmonic wave with index $m=1$ is canceled as expected, whereas the other contributions appear as increased in the overall process. For sake of comparison, also the impedance value $Z_s=-i 339.2\ \Omega$ for the suppression of the harmonic index $m=2$ is reported. 

One can conclude that, when the dimension of the object with respect to the incoming wavelength continues increasing, $N$ impedance coatings are necessary in the radial direction for handling the cancellation of $N$ harmonic indices simultaneously. On the other hand, controlling more than a single dominant harmonic but in the azimuthal direction, it means a metasurface cloak having a surface impedance with oscillating resistance and reactance  values \cite{Sounas}.

In conclusion, a closed-form solution for achieving exact harmonic cancellation has been reported for conductive objects when an impedance metasurface is employed (as the graphene-based one \cite{Dana_apop}). Theoretical predictions are validated in the quasi-static regime and compared with existing solutions \cite{Monti}. Beyond the subwavelength limit, a full Mie analysis \cite{Cui} confirms that the selected scattering coefficient has been canceled. An extension with multilayer coatings, for suppressing more than one dominant harmonic wave in the azimuthal or radial direction, thus going beyond the unidirectionality issue of balanced loss and gain metasurface cloak \cite{Sounas}, is currently under investigation. 

\section*{Acknowledgements}

G. Labate would like to acknowledge all the members of the Postgraduate Office at Heriot-Watt University for useful discussions and suggestions during his visiting research period. In addition, he is grateful to the James Clerk Maxwell foundation for the inspiring visit at Maxwell's birthplace, 14 India Street, Edinburgh.

\end{document}